\documentclass{JHEP3}

\newcommand{\x}{\bm x}
\newcommand{\y}{\bm y}

\newcommand{\z}{\bm z}
\newcommand{\p}{\bm p}
  \usepackage{latexsym,bm,amsmath,amssymb,amsfonts}
  \usepackage{epsfig,graphics,graphicx}
  \usepackage{slashed}
  \usepackage{cite}
  \usepackage[latin1]{inputenc}

  \long\def\comment#1{ }

  \newcommand{\abar}{\bar{\alpha}_s}
  
  \newcommand{\del}{\partial}

  \newcommand{\mcal}{\mathcal}
  
  \newcommand{\beq}{\begin{eqnarray}}
  \newcommand{\eeq}{\end{eqnarray}}
  
 \def\simge{\mathrel{%
   \rlap{\raise 0.511ex \hbox{$>$}}{\lower 0.511ex \hbox{$\sim$}}}}
\def\simle{\mathrel{
   \rlap{\raise 0.511ex \hbox{$<$}}{\lower 0.511ex \hbox{$\sim$}}}}

\vspace{1.5cm}

\preprint{
}

\title{\rm \LARGE Soft gluons away from jets: distribution and correlation}

\author{Emil Avsar\\ Institut de Physique Théorique de Saclay, F-91191 Gif-sur-Yvette, France\\
 E-mail: \email {emil.avsar@cea.fr}}

\author{Yoshitaka Hatta and Toshihiro Matsuo\\Graduate School of Pure and Applied Sciences, University
of Tsukuba, Tsukuba, Ibaraki 305-8571, Japan\\
E-mail: \email
{hatta@het.ph.tsukuba.ac.jp}, \email{tmatsuo@het.ph.tsukuba.ac.jp
 }}

\abstract{ Recently, an exact conformal mapping between soft gluons emitted from jets at large angle
in $e^+e^-$--annihilation and
those in the BFKL evolution of a high energy hadron has been proposed.
We elucidate some remarkable aspects of this
correspondence and use them to analytically compute the distribution and correlation of gluons
 in the interjet region. We also establish the  timelike counterpart of Mueller's
dipole model and discuss the resulting linear and nonlinear evolution equations.

}

\begin{document}

\section{Introduction}
Electron--positron ($e^+e^-$) annihilation into hadrons is one of the most well--studied
high energy reactions that offers a broad arena for testing perturbative QCD predictions
\cite{Kluth:2006bw}. A number of fixed--order and resummed  calculations with ever increasing
 precision have been developed for a variety of observables ranging from jet cross sections to
event shape variables. The impressive agreement between these predictions and experiment  witnessed
 over the past few decades undoubtedly represents a major success of perturbative QCD.

While a large fraction of the theoretical activity in $e^+e^-$--annihilation has been centered
around jet--related observables, there is a great deal of physics to be explored in regions
\emph{between} jets. A primary example is the energy flow $E_{out}$ \cite{Sveshnikov:1995vi,
Korchemsky:1997sy,Oderda:1998en,Belitsky:2001ij,Berger:2001ns},
 the total amount of energy radiated into a specified angular region away from the hard jets. The
underlying partonic process that pertains to  interjet observables is the multiple emission of
soft gluons at wide angle.  In perturbative calculations, large logarithms,
 called `non--global logarithms'  \cite{Dasgupta:2001sh,
Dasgupta:2002bw,Banfi:2002hw,Dokshitzer:2003uw,Berger:2003iw,Marchesini:2004ne,Forshaw:2005sx,
Forshaw:2006fk,Banfi:2006gy}, of the type $(\alpha_s\ln Q/{\mathcal M})^n$ appear,
where $Q$ is the center--of--mass energy and ${\mathcal M}$
is a second hard scale $Q\gg {\mathcal M}\gg \Lambda_{QCD}$ characterizing the observable of interest.
[${\mathcal M}=E_{out}$ in the case of energy flow.] These logarithms arise due to emissions from
secondary gluons and are therefore sensitive to complicated multi--gluon configurations in interjet
regions. Intertwined  with the Sudakov logarithms \cite{Berger:2001ns} associated with emissions from
the primary hard partons, their resummation is a challenging task in perturbation theory (see, however,
\cite{Berger:2003iw}).

 Initially, the resummation of non--global logarithms to all orders was done  numerically, in
Monte Carlo simulations in the large--$N_c$ limit \cite{Dasgupta:2001sh,Dasgupta:2002bw,Banfi:2006gy}.
On the other hand, the authors of  \cite{Banfi:2002hw,Marchesini:2003nh} have succeeded in resumming
logarithms  in the form of evolution equations. Very interestingly, their results bear a striking
resemblance to the BFKL \cite{Kuraev:1977fs,Balitsky:1978ic} and the BK \cite{Balitsky:1995ub,Kovchegov:1999yj}
  equations which have been hitherto discussed exclusively in the context of high energy (Regge) scattering.
 Indeed, the equations in \cite{Banfi:2002hw} and \cite{Marchesini:2003nh}  are almost identical in form to
the BK and BFKL equations respectively, after merely replacing the kernel of the evolution equations as
 \beq
\frac{d^2\Omega_c}{4\pi}\frac{1-\cos \theta_{ab}}{(1-\cos \theta_{ac})
(1-\cos \theta_{cb})} \to \frac{d^2\x_c}{2\pi}\frac{\x^2_{ab}}{\x_{ac}^2\x_{cb}^2}\,.
\label{obs}
\eeq
Here, the left--hand--side is the well--known radiation function of a soft gluon (labeled $c$) from parent
partons $a$ and $b$ ($\theta_{ab}$, etc. are relative angles between momenta.), whereas the right--hand--side
 is the dipole splitting probability in impact parameter space ($\x$'s are two--dimensional vectors and
 $\x_{ab}\equiv \x_a-\x_b$) which is the fundamental building block of the dipole formulation of the BFKL
equation \cite{Mueller:1993rr}.

Such a resemblance naturally prompts one to seek a possible relationship between the two processes
($e^+e^-$ vs. Regge) at a fundamental level.
In \cite{Hatta:2008st}, it was pointed out that the map (\ref{obs}) is a conformal transformation,
known as the stereographic projection.
Moreover, in the strong coupling limit of ${\mathcal N}=4$ supersymmetric Yang--Mills (SYM) theory
 the \emph{same} transformation exactly relates the final state in $e^+e^-$--annihilation and the high energy
hadronic wavefunction in  impact parameter space. The way this latter result was derived (see
\cite{Hatta:2008st} for the details)  emphasizes that in the soft sector the two processes are one
 and the same phenomenon, the only difference being the choice of the coordinate system  in which to
express its physics content. In view of this, the correspondence (\ref{obs}) at weak coupling is hardly
accidental, but must have a deep geometrical origin that goes beyond the  perturbative framework.

The purpose of this paper is to establish a detailed dictionary of the transformation rules  and  explore its  physical consequences.
 The practical advantage in doing so is that on the BFKL side a number of exact analytical results are
known with the help of conformal (or rather, the Möbius) symmetry. Making the most of the dictionary,
one can obtain analytical insights  which incorporate  the effects of the resummation into the partonic
final state in $e^+e^-$--annihilation. We start by in the next section reviewing some basic facts about the stereographic
 projection which realizes the map (\ref{obs}). We clarify how the map correctly accounts for the subtle difference
 in kinematics between the two processes. Then in Section~3, we discuss the single
 and double gluon angular distributions and related observables in the interjet region. In Section~4, we  construct the exact
 timelike analog of Mueller's dipole model using the generating functional technique. In light of this,
 the nonlinear evolution equations in the timelike  and spacelike
 contexts can be treated in a unified fashion. Then in Section~5,
we study the correlation of dipoles (heavy--quark pairs)
in the interjet region based on the results in \cite{Hatta:2007fg,Avsar:2008ph}. Finally in Section~6, we look into the small--angle (distance) limit
of the energy correlation functions in the dipole model and reproduce the OPE result in \cite{Hofman:2008ar}.

\section{Stereographic projection}

The stereographic projection is a mapping between the unit sphere
with coordinates $\Omega=(\theta,\phi)$ and the two--dimensional plane $\x=(x^1,x^2)$.
It is defined by the relations
\beq
x^1=\frac{\sin \theta \cos \phi}{1+\cos \theta}\,,
\quad x^2 =\frac{\sin \theta \sin \phi}{1+\cos \theta}\,, \label{ste}
\eeq
 or equivalently,
 \beq
 \cos \theta=\frac{1-\x^2}{1+\x^2}, \quad \sin \theta=\frac{2|\x|}{1+\x^2},
\quad \cos \phi=\frac{x^1}{|\x|}, \quad
 \sin \phi=\frac{x^2}{|\x|}\, .
\eeq
  The squared length transforms as
\beq
 (d\x)^2=\frac{1}{(1+\cos \theta)^2}(d\theta^2 + \sin^2 \theta d\phi^2)
\equiv \frac{1}{(1+\cos \theta)^2}d\Omega^2\,, \label{ste2}
\eeq
and the area element as
 \begin{align}
 d^2\Omega=(1+\cos \theta)^2 d^2\x =\frac{4}{(1+\x^2)^2}d^2\x \, . \label{area}
 \end{align}
 If one thinks of the sphere as being embedded in a three dimensional
space $(y^1,y^2,y^3)$, the stereographic map can be viewed as a
part of the following conformal transformation in four-dimensions \cite{Cornalba:2007fs};
\begin{align}
x^+=-\frac{1}{2y^+}, \quad x^-=y^--\frac{\y^2}{2y^+}, \quad \x=\frac{\y}{\sqrt{2}y^+}\,,
\label{conf}
\end{align}
where $\y=(y^1,y^2)$ and $x^{\pm}=(x^0\pm x^3)/\sqrt{2}$, $y^\pm=(y^0\pm y^3)/\sqrt{2}$.

We shall regard $\x$ as the transverse plane
perpendicular to the direction of a high energy hadron.\footnote{In the presence of conformal symmetry, it is natural to define $\x$ to be a dimensionless variable. When necessary, one can easily restore the length scale in the problem.   }
 The operator which measures
 the energy density at $\x$ is given by
 \begin{align}
   {\mathcal E}(\x) =\frac{1}{\sqrt{2}}\int_{-\infty}^{\infty} dx^-  T_{--}(x^+=0,x^-,\x)\,.
\label{energy} \end{align}
   On the other hand, the coordinates $\Omega=(\theta,\phi)$ are identified with the polar coordinates
of quarks and gluons in the final state of $e^+e^-$--annihilation.
 The total four--momentum as measured in the $y$--coordinates is related to ${\mathcal E}(\x)$
via the following rules \cite{Hofman:2008ar}
 \beq
 P^+=\sqrt{2}\int d^2\x\, {\mathcal E}(\x),\label{Pplusid}\\
 P^-=\sqrt{2}\int d^2\x \,\x^2 {\mathcal E}(\x),\label{Pminusid} \\
 {\bm P}=2\int d^2\x\, \x {\mathcal E}(\x)\, . \label{third}
 \eeq
 In particular  the energy is given by
 \beq
 E = \int d^2\Omega \,  {\mathcal E}(\Omega)= \frac{1}{\sqrt{2}}(P^++P^-) =
 \int d^2\x (1+\x^2)\, {\mathcal E}(\x)\, ,
 \eeq
  where the operator \cite{Korchemsky:1997sy,Sveshnikov:1995vi}
 \beq
{\mathcal E}(\Omega) \equiv \lim_{r\to \infty}r^2\int_0^{\infty}
dy^0\, n_iT^{0i}(y^0,r\vec{n})= \frac{2}{(1+\cos \theta)^3}{\mathcal E}(\x)\,, \label{kon}
\eeq
  with $\vec{n}=(\sin \theta \cos \phi, \sin \theta \sin \phi, \cos \theta)$
 and $r=\sqrt{y_1^2+y_2^2+y_3^2}\,$,
measures the total energy flowing into the
direction  $\Omega$.  In (\ref{kon}), the factor of 2
accounts for the fact that to a high energy hadron
with four--momentum $(E,0,0,E)$ in the $x$--coordinates corresponds
a  virtual static photon with four--momentum $(Q=2E,0,0,0)$ in the $y$--coordinates.

In the following we shall be often interested in the energy and momentum of individual gluons  rather than the total four--momentum. The transformation rules \eqref{Pplusid}--\eqref{third} instruct us to make the identifications
\beq
\p &\leftrightarrow& 2\x \, k^0\,, \label{11}\\
p^0 &\leftrightarrow& (1+\x^2)k^0\,, \label{22}
\eeq
where we employ the convention that the four--momentum of gluons in the timelike cascades (the $y$--coordinates)
is denoted by $p^\mu$, while that in the spacelike cascades (the $x$--coordinates) is denoted by $k^\mu$.
[In fact, in the present approach the transverse momentum in
the spacelike problem ${\bm k}$ does not appear explicitly, but only implicitly
as the inverse of the transverse coordinates, $\x \sim {\bm k}^{-1}$.]
Note that equations \eqref{11} and \eqref{22} are not independent
of each other due to the identity
\begin{align}
\frac{|\p|}{p^0}=\sin \theta =
\frac{2|\x| }{1+\x^2}\,.
\end{align}

So far, the stereographic projection (\ref{ste}) has been introduced merely as a rule to associate particles living in two different coordinate systems. However, it turns out that this correspondence is preserved by the QCD evolution in the soft approximation. As observed in \cite{Hatta:2008st}, the differential
probability of emitting a soft gluon
 from a dipole (a quark--antiquark pair) with opening angle $\theta_{ab}$
\beq
\abar\frac{d^2\Omega_c}{4\pi} \frac{1-\cos \theta_{ab}}{(1-\cos \theta_{ac})
(1-\cos \theta_{cb})}\equiv \abar d^2\Omega_c K_{ab}(\Omega_c), \,\,\,\,\,\, \abar=\alpha_sN_c/\pi,
\label{time}
\eeq
is exactly mapped via the stereographic projection onto
the gluon emission probability
    from a dipole with transverse size $\x_{ab}= \x_a-\x_b$ in the
spacelike process,
   \beq
 \abar  \frac{d^2\x_c}{2\pi} \frac{\x^2_{ab}}{\x_{ac}^2\x_{cb}^2}\equiv \abar d^2\x_c\, K_{ab}(\x_c)  \,.
\label{tan}
\eeq
[Thus we use the same notation $K$ in both cases but from the argument of $K$
it should always be obvious whether we mean \eqref{time}
or \eqref{tan}.]
This implies that, at least to leading logarithmic accuracy (and in fact in the large--$N_c$ approximation), the high energy (small Bjorken--$x$) QCD evolution in the transverse plane is
 equivalent to the small Feynman--$x$ structure of the interjet parton shower
in $e^+e^-$ annihilation.  In order to genuinely establish this statement and also
to discuss its limitations we must, however, take a closer look into the details of
 kinematics. Indeed, there is a subtle but conceptually important difference between
the two processes which needs to be addressed: When describing parton cascades in $e^+e^-$ annihilation,
it is common  to use the variable $|\p|$ as the evolution parameter: One usually
starts with a large value of $|\p|$ set by the splitting of the photon
into the quark--antiquark pair, and evolves the cascade towards smaller values of $\p$
 with  strong ordering in their magnitude $|\p|$. It is then natural to use
$Y_t \equiv$ ln$(E/|\p|)$ as the evolution ``time''. In contrast,
 in the spacelike parton cascade of a high energy hadron
the transverse momentum is more or less constant while one has strong ordering in energy,
 which in turn implies  strong
ordering in angle. One can then use either $k^0$ or the emission angle $\theta$
as the evolution parameter with the evolution time
$Y_s =\ln (E/k^0)$ or $Y_s= \ln\theta$, and to leading order the two choices
should be equivalent.
 This difference in kinematics is an unavoidable feature of
the multiple soft gluon emission in each case \cite{Marchesini:2003nh}, and might be regarded as an
obstacle against any attempt to find an exact mapping between the
two processes. Remarkably, however, the stereographic
projection automatically converts the nature of the evolution
 parameter into the desired form. Indeed, using (\ref{11}) we have
  \beq
Y_t = \mathrm{ln}\frac{E}{|\p|}  \leftrightarrow
 \mathrm{ln} \frac{E}{2|\x|k^0} \,. \label{arrow}
\eeq
 Due to our identification of $\x$ as the inverse transverse momentum ${\bm k}^{-1}$,
the product $|\x|k^0 \sim k^0/|{\bm k}| =1/\theta$ is indeed a measure of
 the emission angle, and thus we are lead to the identification $Y_t \leftrightarrow Y_s$.
This assures that, to leading logarithmic accuracy, the correspondence works perfectly including the details
 of kinematics, and the whole machinery
developed  for the dipole formulation of the BFKL evolution can be used to
analyze interjet observables in $e^+e^-$ annihilation, or \emph{vice versa}.

 Let us conclude this section with a few additional remarks:

    It seems that the correspondence crucially relies on conformal symmetry, and as such,
it may not hold, or at least needs to be modified, in the next--to--leading logarithmic (NLL)
approximation in QCD where the running coupling effect breaks conformal symmetry. Indeed,
 as already pointed out in \cite{Marchesini:2003nh} the argument of the running coupling
 should be $|\p|$ and $|{\bm k}|$ in the two cases, respectively. This does not agree with the
rule (\ref{11}) derived from a consideration of conformal symmetry alone. On the other hand, in ${\mathcal N}=4$ SYM  theory which is conformal,  there is a good possibility that the correspondence holds to all
orders in the soft approximation, as indicated by the fact that it holds exactly in the strong coupling limit \cite{Hatta:2008st}. The recent NLL result reported in \cite{Balitsky:2008rc} is
very encouraging from this point of view. In Appendix~A we transcribe their result to obtain the timelike NLL dipole kernel.

It is worth mentioning that the collinear singularity $|\p|\sim \theta \to 0$ in
the timelike emission kernel (\ref{time}) maps onto the ultraviolet singularity
$|\x|\sim 1/|{\bm k}| \to 0$ of the spacelike emission kernel (\ref{tan}). Conversely, the
ultraviolet region $|\p| \to \infty$ maps onto the collinear
region $|{\bm k}| \to 0$. Also note that in the timelike case this collinear singularity is
responsible for generating angular--ordered gluons surrounding the primary quark and antiquark
which eventually materialize into observed jets of hadrons with the multiplicity given by the
 standard double--logarithmic formula \cite{Bassetto:1984ik}. These collinear gluons are also
mapped onto the transverse plane of a high energy hadron via the stereographic projection, although
 in this  latter case they are irrelevant  fluctuations which are basically invisible in the scattering process.
In Appendix B, as another interesting aspect of this correspondence, we describe
the interplay between the boost in the $y^\mu$--frame  and
 the  scale transformation (dilatation) in the $x^\mu$--frame originally noted in \cite{Hofman:2008ar}.

\section{Interjet gluon distribution}

\subsection{Single gluon distribution}
As a concrete example of the above mapping, let us compute the
distribution of a single gluon emitted from a color dipole (a $q\bar{q}$
pair). We begin with the $\x$--space and denote by $\x_{a(b)}$ the coordinates of the
 quark (antiquark). The single gluon distribution is
\beq
\frac{d^2N}{d^2\x}
=\abar K_{ab}(\x)\int \frac{dk^0}{k^0}
= \abar K_{ab}(\x)\int^{\ln(E/\Lambda)}_{0} dY_s
\,,
\label{spe}
\eeq
where $Y_s=\ln \frac{E}{2|\x|k^0}$ and $\Lambda$ is a cutoff.\footnote{
 The limits on the $Y_s$ integral would
imply  $1>\theta> \Lambda/E$ in terms of the emission angle $\theta\approx |\bm{k}|/k^0 \sim \Lambda/|\x|k^0$. }
Applying the stereographic projection, we obtain
 \beq
 \frac{d^2N}{d^2\Omega}&=&\frac{1}{(1+\cos \theta)^2} \frac{d^2N}{d^2\x}
 =\abar K_{ab}(\Omega) \int^{\mathrm{ln}(E/\Lambda)}_0 dY_t   \nonumber \\
&=& \frac{\bar{\alpha}_s}{4\pi}\frac{1}{\sin^2\theta}\int_{\Lambda^2}^{E^2} \frac{d\p^2}{\p^2}
\,, \label{lead}
\eeq
 where in the last equality we consider the back--to--back jets configuration $\theta_a=0$,
$\theta_b=\pi$,
which corresponds to the choice
$\x_a=0$, $\x_b=\infty$, see figure \ref{fig1}. Note that we have made the substitution
$Y_s \to Y_t = $ ln$(E/|\p|)$ following
 (\ref{arrow}).
\FIGURE[t]{
\centerline{\epsfig{file=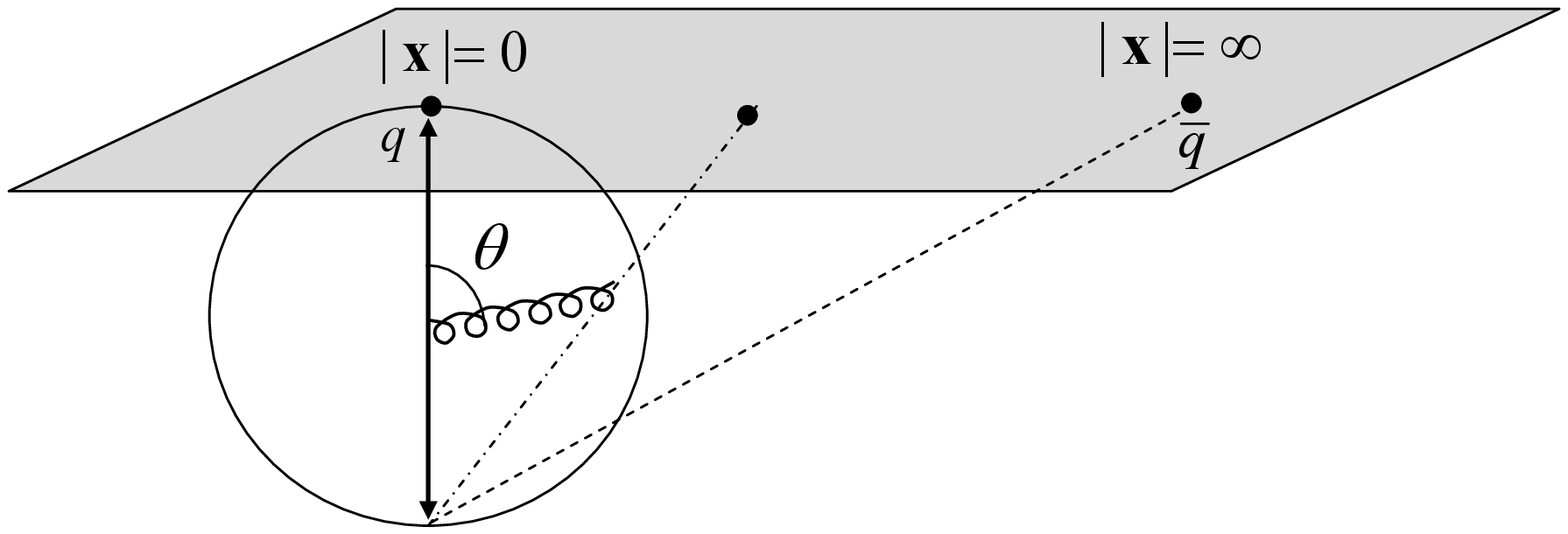,height=4.5cm,width=13.cm}}
\caption{\sl
 Stereographic map between the unit sphere and the transverse plane.
 \label{fig1}}
}

The energy distribution can be similarly computed. In the spacelike case we have
 \beq
 {\mathcal E}(\x)= \abar K_{ab}(\x) \int^{\mathrm{ln}(E/\Lambda)}_{0} k^0\, dY_s  =
  \abar K_{ab}(\x)
\int^{\mathrm{ln}(E/\Lambda)}_{0} \frac{Ee^{-Y_s}}{2|\x|}\, dY_s\,. \label{5}
 \eeq
Using (\ref{kon}), we find, for the back--to--back jets,
\beq
 {\mathcal E}(\Omega) &=& \frac{2}{(1+\cos\theta)^3}{\mathcal E}(\x)
\nonumber \\
&=& \frac{2}{(1+\cos\theta)^3}\frac{\bar{\alpha}_s}{2\pi}
\frac{(1+\cos\theta)^2}{\sin^2\theta} \int  \,
\frac{Ee^{-Y_t}}{2\sin\theta/(1+\cos\theta)} dY_t \nonumber \\
&=& \frac{\bar{\alpha}_s}{4\pi}\frac{1}{\sin^3\theta}
\int d\p^2 \, \frac{|\p|}{\p^2}\,,
\eeq
 in agreement with \cite{Belitsky:2001ij}.
We note, incidentally, that in the $\x$ space the energy distribution
far away from the parent dipole falls as
$\sim 1/|\x|^5$ as can be seen in (\ref{5}).

\subsection{Resummation to all orders}

The results (\ref{spe}) and (\ref{lead}) can be viewed as the leading order term
in the expansion in powers of $\bar{\alpha}_s Y$.\footnote{In the following
we do not distinguish $Y_t$ from $Y_s$.}
If $\bar{\alpha}_s Y$ becomes of order unity, one has to resum all the higher
order terms $(\bar{\alpha}_s Y)^n$ consistently. In the spacelike case, this can be done
 in Mueller's dipole model \cite{Mueller:1993rr,Mueller:1994gb}. In this
approach the average total number of dipoles $(\x_1,\x_2)$ within a rapidity interval $Y$ contained in the parent dipole $(\x_a,\x_b)$
   is given by\footnote{Our normalization of $n_Y$ differs from the usual one in the
literature by a factor of $1/(2\pi \x_{12}^2)$. Note also that despite our somewhat
sloppy notation,  $n_Y$ in fact depends separately on $\x_a$, $\x_b$, $\x_1$ and $\x_2$.
The same remark applies to the other distributions to be defined later. }
 \begin{align} &n_Y(\x_{ab},\x_{12})=\frac{16}{2\pi(\x_{12}^2)^2} \sum_n \int_{-\infty}^\infty
 \frac{d\nu}{(2\pi)^3}
 \left(\nu^2+\frac{n^2}{4}\right)\,  e^{\abar \chi(n,\nu)Y}
  \nonumber \\ & \quad \times \int d^2\x_c
E^{1-h,1-\bar{h}}(\x_{1c},\x_{2c})
E^{h,\bar{h}}(\x_{a c}, \x_{bc})\,,
 \label{bdep}
\end{align}
 where $\chi(n,\nu)$
 is the usual BFKL eigenvalue, and
 $E^{h,\bar{h}}$ is the eigenfunction of the SL(2,${\mathbb C}$) Casimir operator  (see, e.g., \cite{Lipatov:1996ts}) with  $h=\frac{1-n}{2}+i\nu$, $\bar{h}=\frac{1+n}{2}+i\nu$.

  Formally, the total number of dipoles (equal to the total number of gluons plus one) is given by the integration
  \begin{align}
  N_Y=\int d^2\x_1 d^2\x_2 \,n_Y(\x_{ab},\x_{12})\,.
\label{dipdens}
  \end{align}
  As we shall soon see, the integral is actually divergent. Ignoring this fact for the moment, let us analyze its structure from the viewpoint of conformal symmetry. The $d^2{\x_c}$
integration in (\ref{bdep}) gives a function (see below) only of the anharmonic ratio
  \begin{align}
  \rho=\frac{z_{ab}z_{12}}{z_{a1}z_{b2}}=\frac{\left(\tan\frac{\theta_a}{2}e^{i\phi_a}-
  \tan\frac{\theta_b}{2}e^{i\phi_b}\right) \left(\tan\frac{\theta_1}{2}e^{i\phi_1}-
  \tan\frac{\theta_2}{2}e^{i\phi_2}\right)
 }{\left(\tan\frac{\theta_a}{2}e^{i\phi_a}-
  \tan\frac{\theta_1}{2}e^{i\phi_1}\right)\left(\tan\frac{\theta_b}{2}e^{i\phi_b}-
  \tan\frac{\theta_2}{2}e^{i\phi_2}\right)}\,, \label{anharmonic} \end{align}
 and its complex conjugate. [We have here introduced the complex coordinates $z_a=x^1_a+ix^2_a$, etc.]
  The square of this ratio is
  \begin{align}
  |\rho|^2=\frac{\x_{ab}^2\x_{12}^2}{\x_{a1}^2\x_{b2}^2}=\frac{(1-\cos \theta_{ab})
(1-\cos\theta_{12})}{(1-\cos \theta_{a1})(1-\cos\theta_{b2})}\,.
  \end{align}
  Moreover, the integration measure
\begin{align}
 \frac{d^2\x_1 d^2\x_2}{(\x_{12}^2)^2} = \frac{d^2\Omega_1 d^2\Omega_2}{4(1-\cos \theta_{12})^2}\,,
 \end{align}
 is conformally invariant so that  $N_Y$ can be expressed
 in a manifestly conformally invariant way. Keeping only the $n=0$ term which is dominant
 at high energy, we get
 \beq
N_Y &=&\int d^2\x_1 d^2\x_2 \,n_Y(\x_{ab},\x_{12})= \int \frac{d^2\x_1 d^2\x_2}{(\x_{12}^2)^2} f(\rho,\bar{\rho})\,, \nonumber \\
&\equiv &\int d^2\Omega_1 d^2\Omega_2 \,n_Y(\Omega_{ab},\Omega_{12})= \frac{1}{4}\int \frac{d^2\Omega_1 d^2\Omega_2}{(1-\cos \theta_{12})^2} f(\rho,\bar{\rho})\,, \label{mm}
\eeq
 where \cite{Lipatov:1996ts,Navelet:1997xn}
\beq
 f(\rho,\bar{\rho})&=& \int_{-\infty}^\infty d\nu \, e^{\abar\chi(\gamma)Y} \Bigl(
 b_\nu |\rho|^{2-2\gamma}
 \,_2F_1(1-\gamma,1-\gamma,
 2-2\gamma;\rho)\,_2F_1(1-\gamma,1-\gamma,2-2\gamma,\bar{\rho}) \nonumber \\
 && \qquad \qquad \qquad \qquad \qquad  + b^*_{\nu}|\rho|^{2\gamma}
  \,_2F_1(\gamma,\gamma,
 2\gamma;\rho)\,_2F_1(\gamma,\gamma,2\gamma;\bar{\rho})\Bigr)\,,
 \label{lev} \eeq
  with $\gamma\equiv \frac{1}{2}+i\nu = h$,
 and
  \begin{align}
 b_{\nu}\equiv
\frac{\nu 2^{4i\nu}}{2i\pi^3}\frac{\Gamma(\frac{1}{2}-i\nu)\Gamma(1+i\nu)}
 {\Gamma(\frac{1}{2}+i\nu)\Gamma(1-i\nu)}\,.
 \end{align}
 In the second line of (\ref{mm}) we are naturally led to define the timelike analog of
 the single dipole distribution $n_Y(\Omega_{ab},\Omega_{12})$, that is, the
 total number of dipoles with opening angle $\Omega_{12}$ contained in the parent
 dipole $\Omega_{ab}$ within a rapidity interval $Y$. A related distribution
 (integrated over $(\Omega_1+\Omega_2)/2$) was previously introduced in
 \cite{Marchesini:2003nh} in the context of the heavy quark pair production
 in $e^+e^-$--annihilation.  We shall discuss more about this in the next section.

  The generalization of the angular distribution (\ref{lead}) would be
 \begin{align}
 \frac{d^2N_Y}{d^2\Omega_1}=
 \frac{(1+\x_1^2)^2}{4}\frac{d^2N_Y}{d^2\x_1} =
\frac{(1+\x_1^2)^2}{4} \,2\int d^2 \x_2 \,n_Y(\x_{ab},\x_{12})\,.
 \end{align}
If we were to assume the $d^2\x_2$ (or $d^2\Omega_2$) integral to be convergent,
then from dimensional analysis the result would be proportional to
 \begin{align}
 \int d^2 \x_2 \, n_Y(\x_{ab},\x_{12}) \propto \frac{1}{\x_1^2}\,,
 \end{align}
 in the limits $\x_a \to 0$, $\x_b \to \infty$, and therefore,
 \begin{align}
 \frac{d^2N_Y}{d^2\Omega_1} \propto
 \frac{(1+\x^2_1)^2}{4\x^2_1} = \frac{1}{\sin^2 \theta_1}\,.
 \end{align}
This has the same angular dependence $1/\sin^2\theta$ as in the lowest order result \eqref{lead}, and in fact it is the unique possibility consistent with boost invariance  \cite{Belitsky:2001ij}. However, as already mentioned the integral is  divergent due to the  singularity  at $\x_2 = \x_1$
  (or $\Omega_2=\Omega_1$). Indeed, when $|\x_{12}|\sim \theta_{12}$ is very small,
$|\rho|$ is much smaller than unity. We may then approximate as
 $\,_2F_1(...;\rho)\approx 1$ and find
\beq
\frac{d^2N_Y}{d^2\Omega_1} \sim
\int \frac{d\gamma}{i} \int d^2\Omega_2 \frac{e^{\abar\chi(\gamma)Y}}{(1-\cos \theta_{12})^{1+\gamma}}\,.
\eeq
 For small values of $\theta_{12}$, the anomalous dimension $\gamma$ at the
saddle point is between 0 and $\frac{1}{2}$, hence the singularity is not integrable.
It would be interesting to see whether this problem is cured in a more refined treatment
including higher order corrections (in particular, the energy conservation) to the BFKL approximation.

On the other hand, the doubly--differential gluon distribution
is finite and can be identified with the dipole density itself.\footnote{Some care must be
taken  in this identification since the two gluons are not selected randomly but are
constrained such that they are neighboring in the color space. Though we presume that such
a concern is immaterial in the large $N_c$ limit at the level of the two--gluon distribution,
the following results may admit somewhat different interpretations such as the measure of color
flow at large angle. [Note that even the softest gluons carry color.] }
When $|\rho|\ll 1$, a
simple analytical estimate is possible. This encompasses two physical situations (see fig.~\ref{pair})
which are mathematically equivalent:
\FIGURE [t]{
\epsfig{file=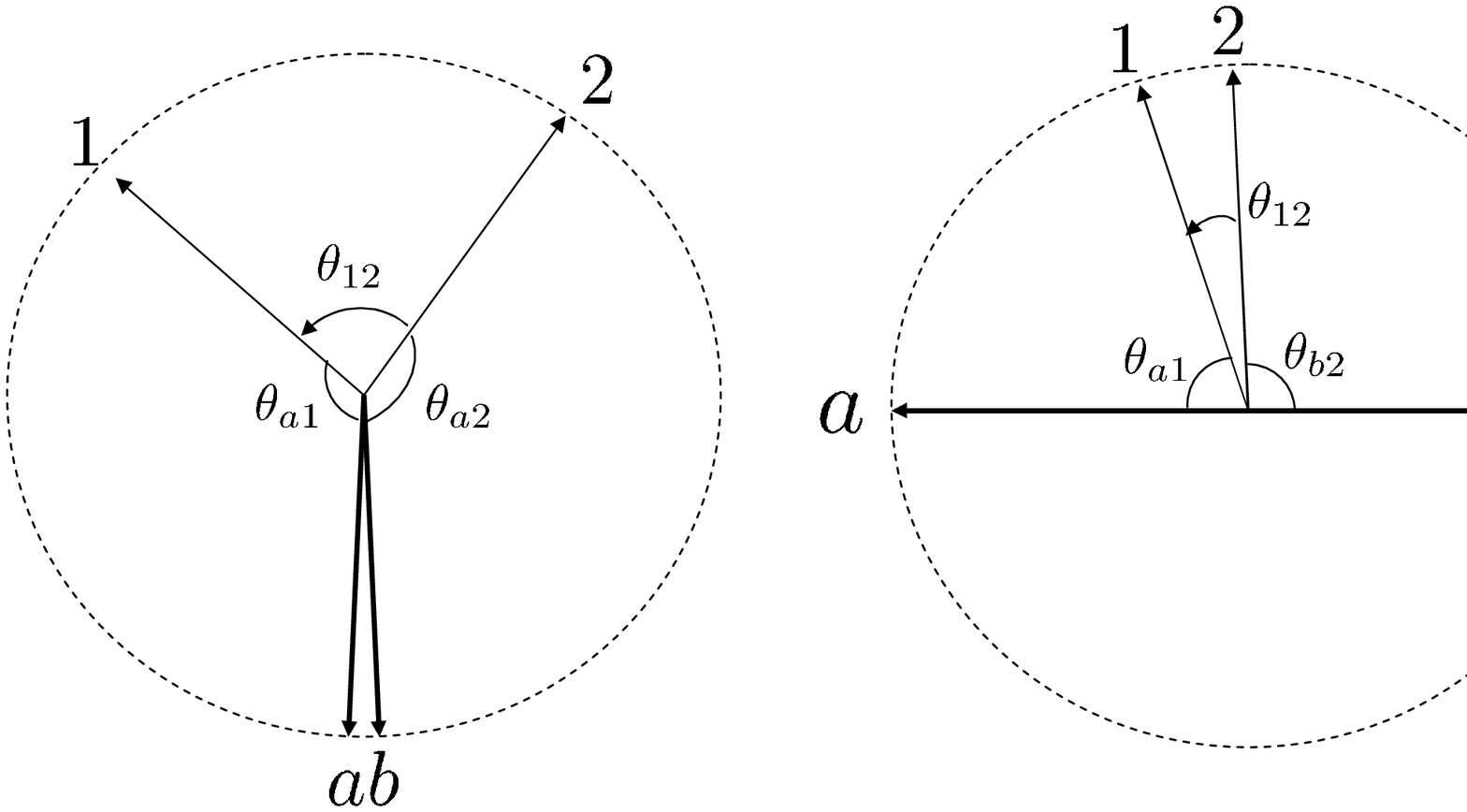,height=6cm,width=11.cm}
\caption{\sl
  The angular correlation of gluons in a highly boosted frame of the $q\bar{q}$ pair (left) and in the center--of--mass frame (right).
  \label{pair}}
}
(i) Radiation from a highly boosted $q\bar{q}$ pair. In this case $\theta_{ab}\ll 1$ and
\beq
|\rho|^2 \approx \frac{\theta_{ab}^2(1-\cos \theta_{12})}{2(1-\cos \theta_{a1})(1-\cos \theta_{a2})} \ll 1\,.
\eeq
Evaluating the $\gamma$ integral in the saddle point approximation, we find
\beq
 \frac{d^4N_Y}{d^2\Omega_1 d^2\Omega_2} \sim \frac{\theta_{ab}^{2-2\gamma_s}\,e^{\abar\chi(\gamma_s)Y}}{(1-\cos \theta_{12})^{1+\gamma_s}(1-\cos \theta_{a1})^{1-\gamma_s}(1-\cos \theta_{a2})^{1-\gamma_s}}\,, \label{approximate}
 \eeq
 where the saddle point $0<\gamma_s<\frac{1}{2}$ is determined by the equation
 \begin{align}
 \chi'(\gamma_s)=-\frac{1}{\abar Y}\ln \frac{1}{|\rho|^2}\,. \label{sp}
 \end{align}
 As an example of the value of $\gamma_s$, we note that $\gamma_s = 0.47$ for $|\rho|^2=0.1$  and  $\abar Y = 2$. As
 $|\rho|^2$ decreases further the saddle point moves slowly towards lower
 values.

  (ii) The back--to-back jets case with $\theta_{12}\ll 1$.
 Setting $\theta_a=0$,  $\theta_b=\pi$ and therefore,
\beq
|\rho|\approx \frac{\theta_{12}}{\sin \theta_{1}} \ll 1\,,  \label{small}
\eeq
we obtain
 \begin{align}
  \frac{d^4N_Y}{d^2\Omega_1 d^2\Omega_2} \sim \frac{e^{\abar \chi(\gamma_s)Y}}
{\theta_{12}^{2+2\gamma_s} (\sin \theta_1)^{2-2\gamma_s}}\,, \label{forward}
  \end{align}
 where the saddle point is again given by (\ref{sp}).

When $\theta_{12}$ becomes large, $|\rho|$ reaches values of order unity. In this
regime  the  hypergeometric function $\,_2F_1(...;\rho)$ has to be fully retained. As an example, let us consider
 the azimuthal correlation of two gluons around the jet axis
by taking $\Omega_a=(0,0)$, $\Omega_b=(\pi,0)$, $\Omega_1=(\frac{\pi}{2},0)$ and $\Omega_2=
(\frac{\pi}{2},\phi)$. Equation (\ref{anharmonic}) becomes
\beq \rho=1-e^{i\phi}\,, \qquad |\rho|^2=2(1-\cos \phi)\,. \eeq
This leads to
\beq
\frac{d^4N_Y}{d^2\Omega_1 d^2\Omega_2}=4\int_{-\infty}^\infty d\nu\frac{b_\nu e^{\abar
\chi(\gamma)Y}}{|\rho|^{2+2\gamma}}
 \,_2F_1(1-\gamma,1-\gamma,
 2-2\gamma;\rho)\,_2F_1(1-\gamma,1-\gamma,2-2\gamma,\bar{\rho})\,. \nonumber \\ \label{www}
\eeq
  We have numerically integrated  the right--hand--side of (\ref{www}). The result is shown
in fig.~\ref{col} as a function of $\phi$ for two different values of $\abar Y$. We see that
the strong correlation in the collinear direction $\theta_{12}=\phi \ll 1$ as described by
(\ref{forward}) decreases towards the backward direction and eventually reaches
a minimum at $\phi=\pi$.
\FIGURE[t]{
\centerline{\epsfig{file=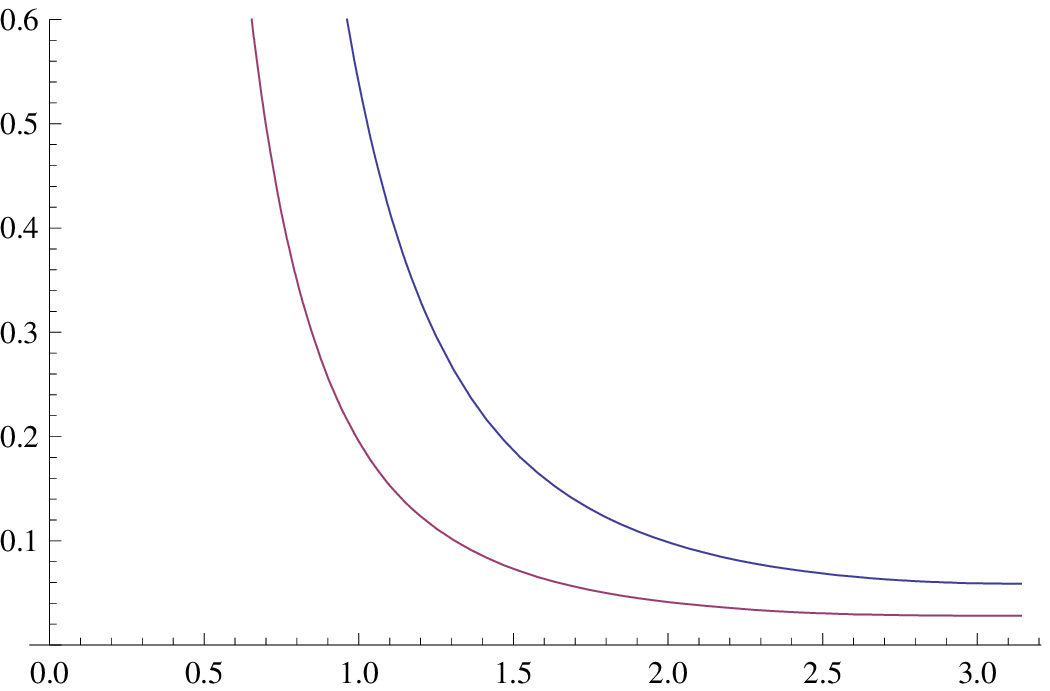,height=4.5cm,width=7.cm}}
\caption{\sl
 The azimuthal  correlation of two gluons around the jet axis for $\abar Y=2$ (upper curve)
and $\abar Y=1.5$ (lower curve). The horizontal axis is $\phi=\theta_{12}$.
 \label{col}}
}

Another interesting physical quantity is the (pseudo--)rapidity correlator of gluons in the interjet region. The rapidity here is defined as
\beq \eta=\ln \cot \frac{\theta}{2}\,, \eeq
in terms of which  the anharmonic ratio reads
\beq
\rho=1-\cot \frac{\theta_1}{2} \tan \frac{\theta_2}{2} e^{i(\phi_2-\phi_1)} = 1-e^{\eta}e^{i\phi}\,.
\eeq
The appearance of the relative rapidity $\eta=\eta_1-\eta_2$ is a consequence of boost invariance \cite{Belitsky:2001ij}.
 Integrating (\ref{www}) over the azimuthal angle, we get\footnote{ It is physically obvious that
the correlator should be invariant under the sign flip $\eta \to -\eta$, thus it is a function only
 of $|\eta|$, or rather, $e^\eta+e^{-\eta}=2\cosh \eta$. In (\ref{numerical}), this can be seen  by
using an identity of the hypergeometric function \beq \,_2F_1(\alpha,\beta,\gamma,z)=(1-z)^{-\alpha}
\,_2F_1\left(\alpha,\gamma-\beta,\gamma, \frac{z}{z-1}\right)\,, \eeq
 or more easily by choosing $\Omega_a=(\pi,0)$, $\Omega_b=(0,0)$.}
\beq
\frac{d^4N_Y}{d\cos \theta_1 d\cos \theta_2}&=&8\pi \int_0^{2\pi} d\phi\int_{-\infty}^\infty
d\nu\frac{b_\nu e^{\abar \chi(\gamma)Y}}{|\rho|^{2+2\gamma}}
 \nonumber \\   &&\times \,_2F_1(1-\gamma,1-\gamma,
 2-2\gamma;\rho)\,_2F_1(1-\gamma,1-\gamma,2-2\gamma,\bar{\rho})\,.  \label{numerical}
\eeq
The large $\eta$ behavior of the above integral can be estimated as follows.
We use the identity
\beq
_2F_1(\alpha,\alpha,2\alpha,\rho) = \frac{2^{2\alpha}}{\sqrt{\pi}\Gamma(\alpha)}\Gamma(\alpha+1/2)
(-\rho)^{-\alpha}Q^0_{\alpha-1}(1-2/\rho), \,\,\,\,\, \rho \notin (0,1),
\eeq
where $Q^\mu_\nu$ is the Legendre function of the second kind.
Since $|\rho|\approx e^{\eta}\gg 1$,
we can neglect the $\rho$ dependence of the Legendre function. [Strictly speaking this
function goes to an infinite constant in the $\rho \to \infty$ limit.]
 Using $\alpha = 1 - \gamma$ in our case, we get
\beq
\frac{d^4N_Y}{d\cos \theta_1 d\cos \theta_2} \sim
\int d\gamma \frac{1}{|\rho|^{2+2\gamma}}\frac{1}
{|\rho|^{2-2\gamma}} I(\gamma, Y) = \frac{1}{|\rho|^{4}} \times I(Y)\,,
\eeq
where $I(Y)$ is the value of the $\eta$--independent integral. We thus see that at large $\eta$ the correlator decays
as $e^{-4\eta}$. Fig.~\ref{rapid} is the result of a numerical integration of (\ref{numerical}) as a
function of $\eta$. The correlator indeed shows an exponential decay
 $e^{-c\eta}$ with $c\approx 4$ already when $\eta \gtrsim 1$,
more or less independently of the value of $\abar Y$.
In comparison, we note that the energy correlation function in the lowest order (two--gluon)
 approximation exhibits a similar exponential decay with $c=3$
 \cite{Belitsky:2001ij}.
\FIGURE [t]{
\epsfig{file=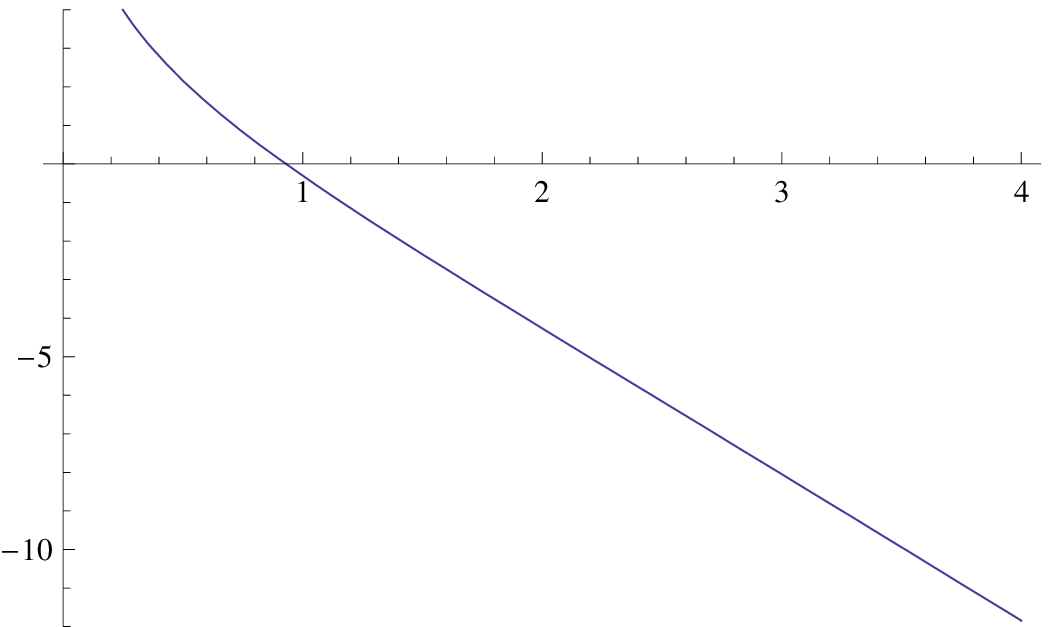,height=6cm,width=9.cm}
\caption{\sl
 Logarithm of the right--hand--side of (\ref{numerical}) for $\abar Y=2$ as a function of $\eta$.
  \label{rapid}}
}

\section{Equivalence of the dipole formulations }
  The single dipole distribution $n_Y(\Omega_{ab},\Omega_{12})$ defined and studied in the previous section is the exact timelike counterpart of the spacelike distribution $n_Y(\x_{ab},\x_{12})$.
In this section we show that this correspondence can be
generalized to the complete equivalence of the dipole formulation of the timelike and spacelike cascades. The generating functional in Mueller's dipole model  is given by
\begin{align}
Z_{ab}[u,Y]=\sum_{n=1}^\infty \int \left(\prod_{i=1}^{n-1} d^2\x_i  \right)
P_n(\x_1,\cdots \x_{n-1};Y)u_1u_2\cdots u_n\,,\label{gen}
\end{align}
where $u_i = u(\x_{i-1},\x_i)$ is an arbitrary `weight' function for the $i$--th dipole and $(\x_0,\x_n)\equiv (\x_a,\x_b)$
is the parent dipole. The function
$P_n$ is the probability distribution to have $n$ dipoles  in a cascade evolved up
to rapidity $Y$. From probability conservation, $Z[u=1] = 1$. The equation satisfied by $P_n$ is
\begin{align}
\partial_Y P_n(\x_1,\cdots \x_{n-1};Y)= -\abar \sum_{i=1}^n
\int d^2\z \, K_{i-1,i}(\z) P_n(\x_1,\cdots, \x_{n-1};Y) \nonumber \\
 + \abar
\sum_{i=1}^{n-1}K_{i-1,i+1}(\x_i)P_{n-1}(\x_1,\cdots
\x_{i-1},\x_{i+1},\cdots \x_{n-1};Y)\,. \label{pps}
\end{align}
This has a simple physical interpretation as a gain--loss type of equation.
The first term on the right hand side is the ``loss'' term and describes the
total probability for the dipole configuration to disappear via all available decay
channels. The second term is the ``gain'' term which is the probability
to obtain the given $n$-dipole configuration from all possible $n-1$ dipole
configurations.
The single dipole density is calculated from $Z_{ab}$ via
\beq
n_Y(\x_{ab},\x_{12}) = \left. \frac{\delta Z_{ab}[u,Y]}{\delta u(\x_1, \x_2)}\right |_{u=1} \,,
\label{density}
\eeq
and it  obeys the BFKL equation
 \beq
\partial_Y n_{ Y}(\x_{ab},\x_{12})&=&\abar\int d^2\x_c \, K_{ab} (\x_c)
\bigl[n_Y(\x_{ac},\x_{12})+n_Y(\x_{cb},\x_{12})-n_Y(\x_{ab},\x_{12})\bigr]
\nonumber \\
&\equiv & \abar\int d^2\x_c \, K_{ab} (\x_c)
\otimes n_Y(\x_{ab},\x_{12})\,.
\label{bfkl}
\eeq
On the other hand, the evolution equation for the generating functional is\footnote{See
\cite{Levin:2003nc} for an illuminating discussion on the consistency between
(\ref{pps}) and (\ref{bk}).}
\beq
\partial_Y Z_{ab}=\abar \int d^2\x_c\, K_{ab} (\x_c)(-Z_{ab}+Z_{ac}\, Z_{cb} )\,.
\label{bk}
\eeq
In particular, if one chooses $u(\x_{i-1},\x_i)=s(\x_{i-1},\x_i)$, the dipole $S$--matrix,
(\ref{bk}) is nothing but the
Balitsky--Kovchegov (BK) equation \cite{Balitsky:1995ub, Kovchegov:1999yj} for the total $S$--matrix $Z_{ab}=S_{ab}$. The
BFKL limit of the BK equation is obtained by defining
the $T$--matrix $t=1-s$, $T=1-S$, and expanding $T$ to linear order in $t$.
 One finds
 \beq
\partial_Y T_{Y}(\x_{ab})&=&\bar{\alpha}_s\int d^2\x_c \, K_{ab}(\x_c)
\bigl[T_Y(\x_{ac})+T_Y(\x_{cb})-T_Y(\x_{ab})\bigr],
 \eeq
 with
 \beq T_Y(\x_{ab})=\int d^2\x_1 d^2 \x_2 \,t(\x_1,\x_2)n_Y(\x_{ab},\x_{12})\,. \label{analog}
 \eeq

 Let us now turn to the timelike cascade and construct the probabilistic interpretation of
the dipole evolution.
First we recall the angular part of the real emission probability of
 $k$ gluons in the soft kinematics \cite{Bassetto:1984ik},
\beq
d{\mathcal P}_k(p_a,p_1,\cdots p_{k},p_b)
= d^2\Omega_1 \cdots d^2\Omega_k
\frac{1-\cos \theta_{ab}}{(1-\cos \theta_{a1})(1-\cos \theta_{12})\cdots (1-\cos \theta_{kb})}\,.
\label{dipweight}
\eeq
In the dipole language this can be viewed as the production probability of
$(k+1)$--dipoles $(a,1), (1, 2), \dots, (k,b)$ obtained by multiplying together successive factors of the kernel $K_{ij}$
according to a particular history of the dipole cascade.
In the final result, the dependence on all the intermediately formed dipoles disappear
and each factor in the denominator associates with one final dipole,
while the numerator comes from the decay of the original dipole $(a,b)$.
Under the stereographic projection, (\ref{dipweight}) becomes
\beq
d{\mathcal P}_k(\x_a,\x_1,\cdots,\x_{k}\x_b)=d^2\x_1\cdots d^2\x_k\frac{\x^2_{ab}}{\x_{a1}^2
\x_{12}^2\cdots\x_{kb}^2}\,,
\label{dipweight2}
\eeq
 which is precisely the corresponding result for the spacelike cascade.
Here the denominator contains a product of the squared length of each
final dipole in the cascade, and the numerator contains the squared length of the
original dipole.

One can similarly show that
the virtual contributions are also mapped onto each other via the
stereographic projection since they are simply obtained by integrating the
real emission kernel $K$ over final state coordinates.
Therefore one is
guaranteed to have the exact timelike analog of the generating functional
techniques described above. This is readily achieved by applying the stereographic projection to  \eqref{gen}
\beq
Z_{ab}[u,Y]&=&\sum_{n=1}^\infty \int \left(\prod_{i=1}^{n-1} \frac{d^2\Omega_i}
{(1+\cos\theta_i)^2} \, \right)
 P_n(\x_1,\cdots \x_{n-1};Y)u_1u_2\cdots u_n \,.
\label{fac}
\eeq
It is natural to define  the probability distributions in the timelike case as
\beq
P_n(\Omega_1,\cdots,\Omega_{n-1};Y) \equiv \frac{ P_n(\x_1,\cdots, \x_{n-1};Y)}
{\prod_{i=1}^{n-1} (1 +\cos \theta_i)^2}
\,. \label{infer}
\eeq
With this definition one can easily check that the evolution equations satisfied by $P_n$ and $Z_{ab}$ in the
timelike case are identical in form to the corresponding equations \eqref{pps} and
\eqref{bk} after replacing $\x_i \to \Omega_i$ everywhere.

 For certain applications, mainly in the timelike context,  the need arises to specify the energy (or rapidity) of individual partons.
 Equation
(\ref{gen}) is not suitable for such purposes since $P_n(...,Y)$ is already integrated
over energy. The more appropriate definition of the generating functional would be\footnote{In fact, this was the original definition employed in
\cite{Mueller:1993rr}.}
\begin{align}
\tilde{Z}_{ab}[u,E] \equiv \sum_{n=1}^\infty \prod_{i=1}^{n} \left(\int d^2\x_i \int^{E}\frac{dk^0_i}{k^0_i}  \right)
\tilde{P}_n(\x_1,\cdots ,\x_{n})u_1u_2\cdots u_{n}\,,
\end{align}
where we have introduced the probability distribution of $n$ gluons (rather than dipoles)
which are ordered in energy, and accordingly, let the source function depend  on one coordinate as well as on  energy
\beq
u_i(\x_{i-1},\x_i) \to u(\x_i,k^0_i)\,.
\eeq
 The equation for $\tilde{Z}$ is \cite{Mueller:1993rr}
\beq
E\partial_E \tilde{Z}_{ab}=\abar \int d^2\x_c \, K_{ab}(\x_c) (-\tilde{Z}_{ab}+u
(\x_c,E)\, \tilde{Z}_{ac}\, \tilde{Z}_{cb} )\,. \label{BMS}
\eeq
In the timelike case, the energy integral becomes $dk^0/k^0=dY_s\to dY_t=dp^0/p^0$ and, due to the correspondence (\ref{dipweight})--(\ref{dipweight2}), it follows that $\prod_i d\x_i P_n(\{\x\}) \to \prod_i d\Omega_i P_n(\{\Omega\})$. Therefore, the equation for $\tilde{Z}$ in the timelike case is identical to (\ref{BMS}) except that $\x_c$ is replaced by $\Omega_c$.

The energy flow observable considered in \cite{Banfi:2002hw} is the probability that
  the total amount of energy emitted into a specified interjet region
${\mathcal C}_{out}$ is less than $E_{out}\ll E$. The nonlinear evolution equation
derived there has precisely the structure (\ref{BMS}) with the weight function
 \beq
u(\Omega_i,p^0_i)= \Theta_{in}(\Omega_i)+e^{-p^0_i/E_{out}}\Theta_{out}(\Omega_i)\,.
\eeq
 This follows from the kinematical constraint
\beq
\Theta(E_{out}-\sum_{j\in {\mathcal C}_{out}} p^0_j)\approx \prod_{i=1}^{n-1}
\left(\Theta_{in}(\Omega_i)+e^{-p^0_i/E_{out}}\Theta_{out}(\Omega_i)\right)\,,
\eeq
where  $\Theta_{out}(\Omega)$ is the support function nonvanishing in ${\mathcal C}_{out}$
and $\Theta_{in}(\Omega) \equiv 1-\Theta_{out}(\Omega)$.

More generally, for each physically motivated choice of $u(\Omega_i,p_i^\mu)$ or $u(\Omega_{i-1},\Omega_i)$
($u(\x_i,k_i^\mu)$ or $u(\x_{i-1},\x_i)$ in the spacelike case)  the generating functional
becomes an observable. Although the interpretations of the resulting evolution
equations (with or without a $u$--factor in the nonlinear term) vary drastically depending
 on the context, mathematically they are equivalent and can be mapped to one another via
the stereographic projection.

\section{Dipole pair density}

Having established the timelike version of Mueller's dipole model, we can now in principle study arbitrary higher order correlations among  dipoles in the interjet region of $e^+e^-$ annihilation.  Physically this is
 relevant to the number correlation of heavy--quark pairs. Similarly to the single dipole distribution $n_Y(\x_{ab},\x_{cd})$ (\ref{density}),
one can define the $k$--dipole inclusive distribution by differentiating $k$ times the generating functional
\beq  n^{(k)}_Y(\x_{ab};\x_{11'},\x_{22'},...,\x_{kk'}) =\frac{1}{k!}\left. \frac{\delta^k Z_{ab}}
{\delta u(\x_{1},\x_{1'})\delta u(\x_2,\x_{2'})
...\delta u(\x_{k},\x_{k'})}\right|_{u=1}\,. \label{kple}\eeq
The corresponding  distribution in the timelike case can be immediately inferred from (\ref{infer}).
In the case of $k=2$ (`the dipole pair density' \cite{Mueller:1994gb}),  we find
\beq  n^{(2)}_Y(\Omega_{ab},\Omega_{11'},\Omega_{22'})= \frac{1}{\prod_{i=11'22'}(1+\cos \theta_i)^2}
\,n^{(2)}_Y(\x_{ab},\x_{11'},\x_{22'})\,. \label{pure} \eeq
 In \cite{Peschanski:1997yx}, an exact integral representation of $n^{(2)}_Y(\x_{ab},\x_{11'},\x_{22'})$
 has been derived. [See also \cite{Braun:1997nu}.] Unfortunately, the expression is too complicated to
be evaluated  in full generality. However, in certain limits analytical results are available
\cite{Hatta:2007fg,Avsar:2008ph}. These include   the large parent limit
\beq |\x_{ab}| \gg |\x_{12}| \gg |\x_{11'}|, |\x_{22'}|\,, \label{AA}  \eeq and the small parent limit
\beq |\x_{a1}|, |\x_{a2}|, |\x_{12}| \gg |\x_{ab}|, |\x_{11'}|, |\x_{22'}|\,. \label{BB} \eeq
 As observed in \cite{Hatta:2007fg}, the above two configurations are transformed to each other via
a conformal transformation. This means that the results can be unified in a single expression which
involves  anharmonic ratios
 \beq
 n^{(2)}_Y(\x_{ab},\x_{11'},\x_{22'}) \sim \frac{e^{2\chi(\gamma'_s)Y}}{\x_{11'}^4\x_{22'}^4}
\left(\frac{\x_{11'}^2\x_{22'}^2}{\x_{12}^4}
 \right)^{1-\gamma'_s}
 \left(\frac{\x_{ab}^2\x_{12}^2}{\x_{a1}^2\x_{b2}^2}\right)^{1-\gamma_s}\,, \label{yo} \eeq
  where the anomalous dimensions $0< \gamma_s< \gamma'_s < 1/2$ are determined from certain  saddle
point conditions.\footnote{See \cite{Hatta:2007fg} for details. Our normalization of $n^{(2)}$ differs
 from that in \cite{Hatta:2007fg} by a factor $(\x_{11'}^2\x_{22'}^2)^{-1}$. Also the anomalous dimension
 is redefined as $\gamma \to 1-\gamma$.} Using (\ref{pure}), (\ref{yo}) and the stereographic projection,
we find
  \beq
  n^{(2)}_Y(\Omega_{ab},\Omega_{11'},\Omega_{22'}) \sim \frac{e^{2\chi(\gamma'_s)Y}}{(1-\cos
\theta_{11'})^2(1-\cos \theta_{22'})^2} \left(\frac{(1-\cos \theta_{11'})(1-\cos \theta_{22'})}
{(1-\cos \theta_{12})^2}\right)^{1-\gamma'_s}
  \nonumber \\ \times \left(\frac{(1-\cos\theta_{ab})(1-\cos \theta_{12})}{(1-\cos \theta_{a1})
(1-\cos \theta_{b2})} \right)^{1-\gamma_s}\,. \label{master}
   \eeq
 The case (\ref{AA}) includes the back--to--back configuration $\theta_{ab}=\pi$. [See fig.~\ref{pair},
 but this time the gluons 1 and 2 are replaced by the dipoles 11' and 22'.] The expression in \eqref{master} then
reduces to
 \beq \label{reduce}
  n^{(2)}_Y(\Omega_{ab},\Omega_{11'},\Omega_{22'}) \sim \frac{e^{2\chi(\gamma'_s)Y}}
{\theta_{11'}^{2+2\gamma'_s} \theta_{22'}^{2+2\gamma'_s}} \frac{1}{(1-\cos \theta_{12})^{1-2\gamma'_s+\gamma_s}}
  \nonumber \\ \times \frac{1}{(1-\cos \theta_{a1})^{1-\gamma_s}(1-\cos \theta_{b2})^{1-\gamma_s}}\,,
   \eeq where the result is valid when $2(1-\cos \theta_{12}) \ll (1-\cos \theta_{a1})(1-\cos \theta_{b2})$.
 On the other hand, the case (\ref{BB}) corresponds to radiation from a highly boosted $q\bar{q}$
jets such that $\theta_{ab}\ll 1$.  Equation \eqref{master} then reduces to
 \beq
  n^{(2)}_Y(\Omega_{ab},\Omega_{11'},\Omega_{22'}) \sim \frac{e^{2\chi(\gamma'_s)Y}}
{\theta_{11'}^{2+2\gamma'_s} \theta_{22'}^{2+2\gamma'_s}} \frac{1}{(1-\cos \theta_{12})^{1-2\gamma'_s+\gamma_s}}
  \nonumber \\ \times \frac{\theta_{ab}^{2-2\gamma_s}}{(1-\cos \theta_{a1})^{1-\gamma_s}
(1-\cos \theta_{a2})^{1-\gamma_s}}\,. \label{with}
   \eeq
 It is interesting to compare (\ref{reduce}) and (\ref{with}) with the two--gluon correlation
 function (\ref{approximate}), (\ref{forward}). Putting aside a possible numerical difference in $\gamma_s$,
 we see that the growth of the correlation  as $\theta_{12}\to 0$ in (\ref{with}) is significantly
weaker than  in (\ref{approximate}). This is of course due to the color screening effect of dipoles.
Moreover, when $\theta_{12}$ becomes comparable to either $\theta_{11'}$ or $\theta_{22'}$ the correlation
function converges to a finite value. This conclusion cannot be reached from (\ref{master})
which assumes $\theta_{12}\gg \theta_{11'}, \theta_{22'}$. Rather, it follows from a  proper
 evaluation of $n^{(2)}_Y$ in the regime $\theta_{12}\approx \theta_{11'}, \theta_{22'}$ (or $|\x_{12}|\approx |\x_{11'}|,|\x_{22'}|$) as was
done in \cite{Avsar:2008ph}. [In the spacelike case this regime deserves special
attention  in association with the BK equation.]

\section{Energy correlation function}

In \cite{Hofman:2008ar}, the small angle limit of the energy--energy
correlation functions in $e^+e^-$ annihilation  was
 studied  using the method of the operator product expansion (OPE)
   \begin{align}
   \langle {\mathcal E}(\Omega_1){\mathcal E}(\Omega_2) \rangle \sim
\frac{1}{|\theta_{12}|^{2+2\gamma(3)}}\,, \quad (\theta_{12}\to 0)\,, \label{in}
    \end{align}
    where $\gamma(j)$ is the anomalous dimension of the twist--two operators with spin $j$.
    Essentially the same OPE applies to the high
energy hadron problem \cite{Hatta:2008st}
 \begin{align}
 \langle {\mathcal E}(\x_1){\mathcal E}(\x_2) \rangle \sim
\frac{1}{|\x_{12}|^{2+2\gamma(3)}}\,, \quad (\x_{12} \to 0) \,. \label{ma}
 \end{align}
 It is straightforward to  generalize these results to higher point correlation functions
 \begin{align}
 \langle {\mathcal E}(\Omega_1){\mathcal E}(\Omega_2)\cdots
{\mathcal E}(\Omega_k)\rangle \sim \frac{1}{\theta^{2k-2+2\gamma(k+1)}}\,,
 \end{align}
 and
 \begin{align}
 \langle {\mathcal E}(\x_1){\mathcal E}(\x_2)\cdots
{\mathcal E}(\x_k)\rangle \sim \frac{1}{|\x|^{2k-2+2\gamma(k+1)}}\,, \label{ii}
 \end{align}
 where we let the relative angles $\theta_{ij}$ and coordinates
$|\x_i-\x_j|$ all go to zero while keeping their ratios fixed, and
denoted their representative values  by $\theta$ and $\x$, respectively.

 The above results are valid not only in  ${\mathcal N}=4$ SYM where they were originally derived \cite{Hofman:2008ar}, but also in any conformal theory. In QCD, they are expected to hold in approximations where the running of the coupling is neglected (as in the leading order BFKL).
In this section we study the energy correlation functions in the dipole model and provide
a  concrete physical picture of how these singular behaviors
are dynamically generated in the parton evolution. While there is the usual caveat about discussing energy--related
observables  in  energy--nonconserving approximations,
  we believe that the dominant configuration which we shall identify below
continues to be the relevant one even after a proper implementation of energy conservation both in the timelike \cite{Lonnblad:1992tz}  and spacelike \cite{Avsar:2005iz} gluon cascades.

\subsection{The energy two--point function}

We first consider  the energy two--point function. Though the primary interest in
 the correlation functions arises in the context of $e^+e^-$ annihilation where they
are experimentally measurable, we find that the actual calculations are a little more
transparent  in the $\x$--space.\footnote{The schematic argument given in \cite{Hatta:2008st}
 overlooks the subtleties to be raised  in the following. Our refined argument is physically
 more correct and readily generalizes to higher order correlation functions.} Of course, via
the stereographic projection all the equations
to follow have a timelike counterpart. We employ the energy ordering scheme and write
 $Y=\ln (E/k^0)$.  The energy two-point function can be generically written as
\begin{align}
\langle {\mathcal E}(\x_1){\mathcal E}(\x_2)
\rangle &=\int_0^\infty dY_2 \int_0^\infty dY_1 \, k^0_1k^0_2 \,
n(\x_1,Y_1,\x_2,Y_2) \nonumber \\ &=2E^2\int_0^\infty
dY_2\, e^{-Y_2}\int_0^{Y_2} dY_1\, e^{-Y_1} n(\x_1,Y_1,\x_2,Y_2) \,, \label{energycorr}
\end{align}
where  $n(x_1,Y_1,\x_2,Y_2)$
is  the number density of gluon pairs
with rapidities $Y_1$ and $Y_2$ located at $\x_1$
 and $\x_2$, respectively. At high energy, we would like to relate
this quantity to the single dipole density $n_Y(\x_{ab}, \x_{12})$. [Henceforth we suppress the
dependence on the parent dipole coordinates $\x_{ab}$ in $n_Y$ and set $|\x_{ab}|=1$.]

In general, there is no universal relation between $n(\x_1,Y_1,\x_2,Y_2)$
and $n_Y(\x_{12})$; the former carries information on  the rapidity of the gluons 1 and 2 (in a cascaded evolved up to $Y\to \infty$), whereas the latter
   simply counts the total number
of dipoles $(\x_1,\x_2)$ in a cascade evolved up to  $Y$ without specifying the
energy of their constituents.  However, as long as  the singular behavior in the limit
$\x_1 \to \x_2$ is concerned, the two quantities can be linked by the following
argument:  For a given gluon pair
with rapidities $Y_1$ and $Y_2$ such that $Y_2>Y_1$, the strongest correlation comes
from the moment
in the history of the  cascade when the two gluons in question
were realized as a dipole as a result of the emission of the gluon with  rapidity $Y_2$.
The number of such pairs at the time of creation is related to $\del_{Y_2}
n_{Y_2}( \x_{12})$ where the derivative is because the gluon 2 has rapidity exactly equal
to $Y_2$ (rather than integrated over rapidity).
This is, however, still inclusive in the rapidity of the gluon 1.
Therefore, we are led to define the ``unintegrated'' dipole density, $n^u(\x_1,Y_1,\x_2,Y_2)$,
as follows
\beq
\del_{Y_2}n_{Y_2}( \x_{12}) = \int_0^{Y_2} dY_1 \,
  n^u(\x_1,Y_1,\x_2,Y_2)\,.
  \label{eq:unintdens}
 \eeq

 To proceed, we observe that $n^u$ is approximately independent of $Y_2$.
In order to see this,
consider a simple toy model of the dipole cascade in which we neglect the transverse
dimensions and  assume each dipole to split after a fixed
rapidity interval\footnote{In the full model one  has
\beq
  \Delta Y  \sim \frac{1}{\abar \mathrm{ln}(\x_{ab}^2/\rho^2)}\,,
  \eeq
  where $\rho$ is a small cutoff.} $\Delta Y$.
  At $Y=0$, there is only one dipole composed of a quark and an antiquark both
having $Y = 0$ which we denote as  $(0,0)$.
  This dipole emits a gluon at rapidity $\Delta Y$ and splits into two dipoles
  $(0,\Delta Y)$ and $(\Delta Y, 0)$.  In the second step these two child dipoles split into
four dipoles $(0,2\Delta Y)$, $(2\Delta Y, \Delta Y)$, $(\Delta Y, 2\Delta Y)$ and ($2\Delta Y, 0)$.
  The process stops
  after $n=Y_{tot}/\Delta Y$ steps, with $Y_{tot}$ being the total rapidity, and generates
$2^n=2^{Y_{tot}/\Delta Y}=\exp\{(\ln 2/\Delta Y)Y_{tot}\}$ dipoles. One can easily see that there are
$2^{(Y/\Delta Y)-1}$ gluons having rapidity $Y$. In particular, half of the total number of gluons
have  the maximal value $Y=Y_{tot}$.
 This means that all the dipoles have the rapidity composition either $(Y_{tot},Y)$ or ($Y,Y_{tot})$ and their
number depends only on $Y$, and not on $Y_{tot}$.
In the full  model the rapidity distribution is  more smeared out than in the
toy model, but it will again
be true that the density $n^u(\x_1,Y_1,\x_2,Y_2) $ is largely determined
by the gluons at $Y_1$, since their number is exponentially smaller than the number of gluons
at $Y_2$ in the presence of the strong energy ordering $Y_2 \gg  Y_1$.

We thus write $n^u=n^u(\x_1, \x_2, Y_1)$  and get
\beq
\del^2_{Y_1}n_{Y_1}( \x_{12}) =
  n^u(\x_1, \x_2, Y_1) \,, \label{hen}
\eeq
 as the number density of dipoles ($\x_1,\x_2$) having the rapidity composition
$(Y_1, Y_2)$ in a cascade evolved up to $Y_2$.
Later in the evolution (i.e., at $Y>Y_2$) these dipoles may or may not split, but the gluons 1
and 2 remain fixed at their original positions $(\x_1,\x_2)$. Since the total
probability of all the decay channels is unity, the most singular contribution (as $\x_{12}\to 0$) to the two--gluon distribution
$n(\x_1,Y_1,\x_2,Y_2)$ in (\ref{energycorr}) which is in principle defined in a cascade
 evolved up to $Y=\infty$ is actually frozen at $Y=Y_2$ (or at whatever value $Y>Y_1$) and is given by (\ref{hen}).

We now recall  the integral expression (\ref{mm}) for $n_Y(\x_{12})$  in the
limit $\x_{12}\to 0$
 \begin{align}
 n_Y(\x_{12})= \frac{1}{\x^2_{12}}\int \frac{dj}{2\pi i}c(j) \frac{e^{(j-1)Y}}
{|\x_{12}|^{2\gamma(j)}}\,, \label{c}
 \end{align}
 where the contour $j$--integral goes around the branch cut in the Jacobian $c(j)$
\begin{align}
c(j)\propto \frac{\partial \gamma(j)}{\partial j}\approx
\frac{-1}{2\sqrt{14\bar{\alpha}_s\zeta(3)(j-j_0)}}\,. \quad (j_0=1+4\bar{\alpha}_s\ln2)
\end{align}
 Using this, we find the energy--energy correlator
\beq
\langle {\mathcal E}(\x_1){\mathcal E}(\x_2) \rangle &\approx &2E^2
\int_0^\infty dY_2 e^{-Y_2}\int_0^{Y_2} dY_1 e^{-Y_1} \del^2_{Y_1} n_{\tiny{Y}_1}(\x_{12})
\nonumber \\
&=&2E^2\frac{1}{\x^2_{12}}\int\frac{dj}
{2\pi i}c(j)\frac{(j-1)^2}{3-j} \frac{1}{|\x_{12}|^{2\gamma(j)}}\,. \label{cut}
\eeq
 Deforming the contour and picking up the pole at $j=3$, we find
\beq
\langle {\mathcal E}(\x_1){\mathcal E}(\x_2) \rangle \sim
\frac{E^2}{|\x_{12}|^{2+2\gamma(3)}}+\frac{1}{\x_{12}^2}
\int_{Re j>3} dj \cdots \frac{1}{|\x_{12}|^{2\gamma(j)}}
 \,.
\eeq
 Since $\gamma(j) < \gamma(3)$ for $j>3$, the contribution from the second
term is subdominant in the limit $\x_{12} \to 0$. We therefore recover the OPE result (\ref{ma}).

\subsection{The energy three--point function and beyond}

The energy three--point function is given by
\begin{align}
\langle {\mathcal E}(\x_1){\mathcal E}(\x_2){\mathcal E}(\x_3)
\rangle =3!E^3\int_0^\infty dY_3 e^{-Y_3}\int_0^{Y_3} dY_2 e^{-Y_2}\int_0^{Y_2} dY_1
 e^{-Y_1} n(\x_1,Y_1,\x_2,Y_2,\x_3,Y_3)  \,, \label{three}
\end{align}
where we  $n(...,\x_3,Y_3)$ is the distribution function of three gluons.
The leading singular behavior as $\x_{ij}\to 0$ arises from the configuration where
the three gluons are realized as two contiguous dipoles which, for $Y_3 > Y_2 > Y_1$, are
$(\x_1,\x_3)$ and $(\x_3,\x_2)$. This occurs when the gluon 3 is emitted from dipoles ($\x_1,\x_2$) whose  number density is given by (\ref{hen}).
Thus to get the number density of the dipole pairs ($\x_1,\x_3$) and ($\x_3,\x_2$)
with rapidity composition $(Y_1, Y_3)$ and $(Y_3, Y_2)$, we just need to multiply (\ref{hen}) by the emission probability. This gives the unintegrated density
\beq
n^u(\x_1,Y_1,\x_2,Y_2,\x_3,Y_3) \approx \abar K_{12}(\x_3) \del^2_{Y_1}n_{Y_1}( \x_{12}) \,.
\eeq

We now substitute this quantity in \eqref{three} as a dipole model analog of
the three gluon distribution. We remind the reader that this is justified only
for the singular contribution in the limit $\x_{ij}\to 0$.
Writing $\x_{ij}\sim \x$ and therefore $K_{12}(\x_3) \sim
\frac{1}{\x^2}$,  \eqref{three} then becomes
\beq
\langle {\mathcal E}(\x_1){\mathcal E}(\x_2){\mathcal E}(\x_3) \rangle
&\sim & E^3\int_0^\infty dY_3 e^{-Y_3}\int_0^{Y_3} dY_2 e^{-Y_2}\int_0^{Y_2} dY_1
e^{-Y_1}   \nonumber \\
&& \times \frac{\abar}{\x^4}\int \frac{dj}{2\pi i}c(j)(j-1)^2e^{(j-1)Y_1} \left(\frac{1}{\x}
  \right)^{2\gamma(j)}\nonumber \\&=&\frac{\abar E^3}{ \x^4} \int \frac{dj}{2\pi i}
c(j)\frac{(j-1)^2}{2(4-j)} \frac{1}{|\x|^{2\gamma(j)}}  \nonumber \\
  &\sim&\abar E^3 \frac{1}{|\x|^{4+2\gamma(4)}}\,,
\eeq
 in agreement with (\ref{ii}).

It is now straightforward to extend the above result to the energy $k$--point correlation function
\beq
\langle \mcal{E}(\x_1) \mcal{E}(\x_2) \cdots \mcal{E}(\x_k)\rangle =
k!E^k\int_0^\infty dY_k e^{-Y_k} \cdots \int_0^{Y_2} dY_1
 e^{-Y_1} n(\x_1,Y_1,\x_2,Y_2, \dots, \x_k,Y_k)\,.  \nonumber \\
 \label{eq:npoint}
 \eeq
The leading singular contribution to the $k$--gluon distribution $n(...,\x_k,Y_k)$ with
$Y_k>Y_{k-1}>\cdots >Y_1$ comes from the process where we start with the lowest rapidity
pair $(\x_1,\x_2)$
and successively emit the remaining gluons, creating first the pair
$(\x_1,\x_3)$ and $(\x_3,\x_2)$, then emitting the gluon 4 from either of these two
dipoles, and so on. One then gets the
fully unintegrated distribution\footnote{This expression is valid for a particular color ordering of the dipoles, but since in the end
we let all $|\x_i-\x_j|$ go to zero with fixed ratios, the result holds for any ordering.}
\beq
\abar^{k-2} K_{12}(\x_3) K_{32}(\x_4) \cdots
 K_{k-1,2}(\x_k)\,\partial_{Y_1}^2 n_{Y_1}(\x_{12})
\sim \left(\frac{\abar}{\x^{2}}\right)^{k-2}
 \partial_{Y_1}^2 n_{Y_1}(\x_{12})\,.
\eeq
Substituting this in (\ref{eq:npoint})  as a proxy for the $k$--gluon distribution $n(...,\x_k,Y_k)$,
one finds the result
\beq
\langle \mcal{E}(\x_1) \mcal{E}(\x_2) \cdots \mcal{E}(\x_k)\rangle &\sim&
\abar^{k-2}E^k \frac{1}{\x^{2(k-2)}}
\int_0^\infty dY_k e^{-Y_k} \cdots \int_0^{Y_2} dY_1 e^{-Y_1} \nonumber \\
&& \times \frac{1}{\x^2}\int \frac{dj}{2\pi i}c(j)(j-1)^2\,e^{(j-1)Y_1} \frac{1}{|\x|^{2\gamma(j)}} \nonumber \\
  &\sim&  E^k \abar^{k-2} \frac{1}{|\x|^{2\gamma(k+1)+2k-2}}\,.
\eeq

 Summarizing, in the dipole picture the most singular contribution to the
$k$--point energy correlation function comes from the moment in the history
of the dipole cascade when the $k$ gluons are realized as a single chain of $k-1$
dipoles. This chain represents successive emissions of energy ordered gluons at
comparable separations (collinear emissions at comparable angles in the timelike case).
 Later in the evolution the chain decays, but the $k$ gluons remain in their original
positions. Integrating over all possible values of rapidity at which
this chain is formed, one recovers the OPE result.

In strongly coupled   ${\mathcal N}=4$ SYM the AdS/CFT correspondence allows one to represent the energy correlation function as the
 scattering amplitude between the photon (the primary $q\bar{q}$ pair) and the graviton (energy operator insertions, or calorimeters) \cite{Hofman:2008ar}. Its small angle limit is  associated with the Regge behavior of the string S--matrix and is dominated by the $t$--channel exchange of a massive string state (`Pomeron')  with an anomalous dimension $\gamma(1+k)$.
A similar interpretation is possible in QCD by identifying the $t$--channel object with the BFKL Pomeron. 
Naively, one may expect from the relevant configurations that the $k$--point function
would be related to the $(k-1)$--ple dipole density $n_Y^{(k-1)}$ (cf. (\ref{kple})) which
 involves the exchange of $(k-1)$ Pomerons with an anomalous dimension
$\gamma=\chi^{-1}(4(k-1)\ln2)$ \cite{Xiao:2007te}. However, specifying the rapidity of
$k$ gluons amounts to taking the $k$--th derivative of $n^{(k-1)}_Y$ with respect to $Y$.
Each $Y$--derivative lowers the degree of multiplicity, thus one has a cascade of
distributions $n^{(k-1)} \to n^{(k-2)} \to n^{(k-3)} \to \cdots$ \cite{Levin:2004yd}. The
fully unintegrated distribution obtained in this way involves the single dipole density
or the single Pomeron exchange with an unusual value of the anomalous dimension $\gamma(1+k)$.

\section*{Acknowledgements}
The work of Y.~H. and T.~M. is supported, in part, by Special Coordination Funds for Promoting Science and Technology of the Ministry of Education, Culture, Sports, Science and Technology, the Japanese Government.

\appendix

\section{Next-to-leading order dipole kernel in $e^+e^-$--annihilation}

 Recently,  Balitsky and Chirilli have derived the spacelike  next--to--leading logarithmic (NLL) dipole
kernel in ${\mathcal N}=4$ SYM which turned out to be conformally (Möbius) symmetric \cite{Balitsky:2008rc} (see, also, \cite{Fadin:2007xy}). As we noted in Section~2, in this theory there is a rather strong indication that the stereographic
projection works to all orders of the soft approximation. Assuming this to be
correct, one can immediately obtain from their result the NLL kernel for the
dipole evolution in the timelike case by applying the stereographic projection. The evolution equation for the dipole
density which generalizes the leading order equation (\ref{bfkl}) is
\beq
\partial_Y n_{ Y}(\Omega_{ab})&=&\abar \left(1-\abar \frac{\pi^2}{12}\right)\int
d^2\Omega_c \, K_{ab} (\Omega_c)
\bigl[n_Y(\Omega_{ac})+n_Y(\Omega_{cb})-n_Y(\Omega_{ab})\bigr]
\nonumber \\
&& \qquad + \abar^2 \int d^2\Omega_c d^2\Omega_d K'_{ab}(\Omega_c,\Omega_d)n_Y(\Omega_{cd})\,,
\eeq
where
\beq
K'_{ab}(\Omega_c,\Omega_d)
&=& \frac{1}{8\pi^2} \biggl\{ \frac{(1-\cos \theta_{ab})}{(1-\cos \theta_{ac})(1-\cos
\theta_{cd})(1-\cos \theta_{db})} \nonumber \\
&&\times \biggl[\left(1+\frac{(1-\cos \theta_{ab})(1-\cos \theta_{cd})}{(1-\cos \theta_{ac})
(1-\cos \theta_{bd})-(1-\cos \theta_{ad})(1-\cos \theta_{bc})}\right) \nonumber \\ && \times
\ln \frac{(1-\cos \theta_{ac})(1-\cos \theta_{bd})}{(1-\cos \theta_{ad})(1-\cos \theta_{bc})}
+2\ln \frac{(1-\cos \theta_{ab})(1-\cos \theta_{cd})}{(1-\cos \theta_{ad})(1-\cos \theta_{bc})}
 \biggr] \nonumber \\  &+ &12\pi^2 \zeta(3)\delta^{(2)}(\Omega_{ac})\delta^{(2)}(\Omega_{bd}) \biggr\}\,.
\eeq
The eigenfunctions and the eigenvalues can be exactly mapped as well.

\section{The correspondence between boost and dilatation}

The correspondence of the gluon distribution has an interesting property under boost.
One can see from (\ref{conf}) that the dilatation in the $x$--coordinates,
\begin{align}
x^\mu \to x'^\mu=
\lambda x^\mu   \label{boost}
\end{align}
 translates into the boost in the $y^3$ direction
\begin{align} y^+ = \lambda y'^+, \quad y^-=\frac{y'^-}{\lambda}, \quad \y = \y'\,.   \end{align}
This transformation relates the gluon distribution in the boosted frame to that generated
by a squeezed  dipole. For simplicity, assume that the initial quark--antiquark pair is
oriented to the $y^1$--axis (see fig.~\ref{figapp}).
\FIGURE[t]{
\centerline{\epsfig{file=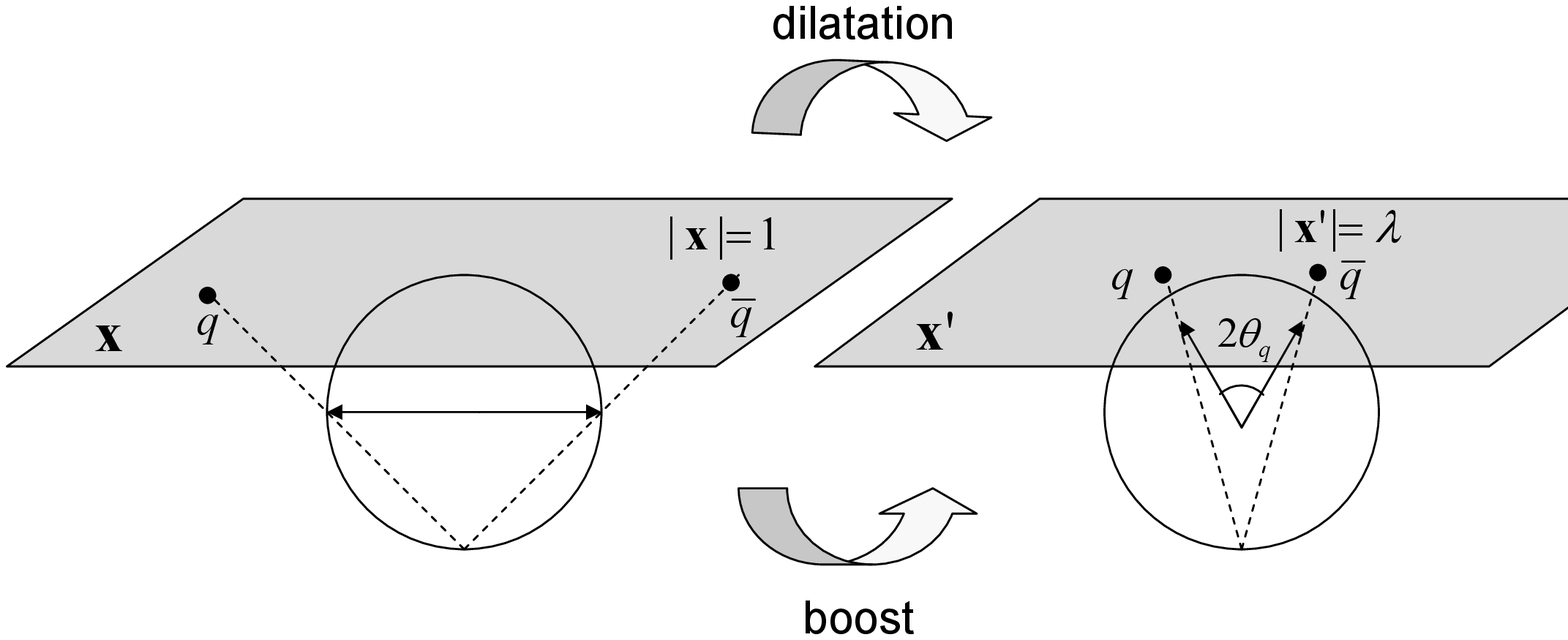,height=6cm,width=14.cm}}
\caption{\sl
 Boosting the $q\bar{q}$ pair in $e^+e^-$ annihilation corresponds to changing the size of the  dipole in
 high energy scattering.
 \label{figapp}}
}
 $\lambda$ is related to the velocity $v$ of the boost via
 \begin{align}
 \lambda =\frac{1}{\gamma (1+v)}\,, \end{align} where $\gamma=1/\sqrt{1-v^2}$ as usual.
 Since $v=\cos \theta_q$ in the new frame where $\theta_q$ is the angle of the quark jet, one has
 \begin{align}
 \lambda=\frac{1}{\gamma (1+v)}= \frac{\sin\theta_q}{1+\cos \theta_q} = |\x|\,, \end{align}
  consistently with (\ref{boost}).

 Under the dilatation (\ref{boost}), the energy distribution (\ref{energy}) transforms as
 \begin{align}
  {\mathcal E}'(\x')=\frac{1}{\lambda^3} {\mathcal E}(\x) \end{align}
 The distribution in the boosted frame is
 \begin{align}
 {\mathcal E}(\Omega') = \frac{2}{(1+\cos \theta')^3}{\mathcal E}(\x')
 = \frac{1}{\lambda^3} \frac{(1+\cos \theta)^3}{(1+\cos \theta')^3} {\mathcal E}(\Omega)
  \end{align}
 Using the relation
 \begin{align}
 \cos \theta'=\frac{v+ \cos \theta}{1+v\cos \theta}
 \end{align}
  one finds
  \begin{align}
  {\mathcal E}(\Omega')=\frac{1}{\gamma^3(1-v\cos \theta')^3} {\mathcal E}[\Omega(\Omega')]
  \end{align}
  Similarly the number (or the charge) distribution transforms as
  \begin{align}
  \frac{dN}{d\Omega'}=\frac{1}{\gamma^2(1-v\cos \theta')^2}  \frac{dN[\Omega(\Omega')]}{d\Omega}
  \end{align}

\bibliographystyle{utcaps}
\bibliography{references}

\end{document}